\documentclass[%
reprint,
 amsmath,amssymb,
 aps,
]{revtex4-1}

\usepackage{color}
\usepackage{graphicx}% Include figure files
\usepackage{dcolumn}% Align table columns on decimal point
\usepackage{bm}% bold math
\usepackage{hyperref}% add hypertext capabilities
\def\Vec#1{\boldsymbol #1}

\newcommand {\beq}{\begin{eqnarray}}
\newcommand {\eeq}{\end{eqnarray}}

\makeatletter
\newcommand{\oset}[2]{%
  {\mathop{#2}\limits^{\vbox to -.5\ex@{\kern-\tw@\ex@
   \hbox{\scriptsize #1}\vss}}}}
\makeatother

\begin{document}

%\preprint{APS/123-QED}

\title{
Berezinskii-Kosterlitz-Thouless Transition of Two-component Bose Mixtures\\ 
with Intercomponent Josephson Coupling
}% Force line breaks with \\
%\thanks{Geometrical determination of internal degrees of freedom localized in vortex cores}%

\author{
Michikazu Kobayashi$^1$,
Minoru Eto$^{2,3}$,
Muneto Nitta$^3$
}
\affiliation{%
$^1$Department of Physics, Kyoto University, Oiwake-cho, Kitashirakawa, Sakyo-ku, Kyoto 606-8502, Japan, \\
$^2$Department of Physics, Yamagata University, Kojirakawa-machi 1-4-12, Yamagata, Yamagata 990-8560, Japan, \\
$^3$Department of Physics, and Research and Education Center for Natural Sciences, Keio University, Hiyoshi 4-1-1, Yokohama, Kanagawa 223-8521, Japan
}%

\date{\today}% It is always \today, today,

\begin{abstract}
We study the Berezinskii-Kosterlitz-Thouless (BKT) transition of two-component Bose mixtures in two spatial dimensions.
When phases of both components are decoupled, half-quantized vortex-antivortex pairs of each component induce two-step BKT transitions. %regardless of the inter-component interaction.
%In this case, we find an unconventional universal relation between the superfluid density and the BKT transition temperature.
On the other hand, when phases of both components are synchronized through the intercomponent Josephson coupling, two species of vortices of each component are bound to form a molecule, and, in this case, we find that there is only one BKT transition by molecule-antimolecule pairs.
Our results can be tested by two weakly connected Bose systems such as two-component ultracold diluted Bose mixtures with the Rabi oscillation, and multiband superconductors.
\end{abstract}

\pacs{}

\maketitle

%\section{Intruduction}

Phase transitions in two-dimensional systems with a continuous symmetry have long attracted much attention since the theoretical prediction by Berezinskii, Kosterlitz, and Thouless (BKT), providing a topological ordering through the binding of vortex-antivortex pairs \cite{Berezinskii,Kosterlitz}.
Being different from the conventional thermodynamic transition prohibited by the Coleman-Mermin-Wagner theorem in two-dimensional systems \cite{Coleman,Mermin,Hohenberg}, the BKT transition exhibits a critical line below the BKT transition temperature, $T \leq T_{\rm BKT}$, with continuously variable critical exponents and the nonzero helicity modulus (superfluid density) showing discontinuous jump at the BKT transition temperature.
The BKT transition has been observed in $^4$He films \cite{Bishop}, thin superconductors \cite{Gubser,Hebard,Voss,Wolf,Epstein}, Josephson-junction arrays \cite{Resnick,Voss2}, colloidal crystals \cite{Halperin,Young,Zahn}, and ultracold atomic Bose gases \cite{Hadzibabic}.

One of important issues of the BKT transition is a relationship between its universality and topological aspects of vortices.
In the two-dimensional Bose systems with no internal degree of freedom, circulations of vortices are quantized by $2 \pi \hbar / m$ with the particle mass $m$, giving the universal jump of the superfluid number density $\Delta \rho_{\rm s}$ at the BKT transition temperature $T_{\rm BKT}$ as
\begin{align}
\Delta \rho_{\rm s} = \frac{2 m T_{\rm BKT}}{\pi \hbar^2}.
\label{eq:single-component-universal-relation}
\end{align}
%One example is many particle system with short-range interactions.
%In three-dimensional systems, a transition occurs from fluid to crystal with spontaneous and simultaneous breakings of translational and rotational symmetries at one transition temperature, giving positional and directional orders.
%These two symmetry breakings allow two types of topological defects, i.e., dislocations and disclinations.
%The fluid-crystal transitions in three-dimensional systems is replaced with two-step transitions at different temperatures in two-dimensional systems with bindings of dislocation-antidislocation and disclination-antidisclination pairs.
%Another example is the two-dimensional system containing vortices classified by nontrivial discrete groups.
%In the case of vortices classified by the finite cyclic group $\mathbb{Z}_2$ which appear in Heisenberg antiferromagnets on the triangular lattice, a vortex becomes its antivortex.
%It remains a big open problem about whether two-dimensional systems containing $\mathbb{Z}_2$ vortices can show the transition for bindings of vortex-(anti)vortex pairs and give the nontrivial nonzero quantity such as the helicity modulus or not.
On the other hand, multicomponent systems in general allow 
quantized vortices with fractional circulations, 
which are studied in superfluid $^3$He \cite{Salomaa,Volovik,Autti}, $p$-wave superconductors \cite{Salomaa,Ivanov,Chung,Jang}, multiband or multicomponent superconductors \cite{Babaev,Goryo,fractional-exp,Tanaka}, spinor Bose systems \cite{Ho,Semenoff}, multicomponent Bose systems \cite{Son,Mueller,Kasamatsu-1,Kasamatsu-2,Aftalion,Kuopanportti,Kasamatsu-3,Kasamatsu-4,Eto:2011wp,Eto-2,Eto:2013spa,Cipriani,Cipriani:2013wia,Dantas:2015fka,Tylutki:2016mgy,Eto:2017rfr,Kasamatsu:2015cia,Uranga}, exciton-polariton condensates \cite{Rubo,Keeling}, nonlinear optics \cite{Pismen}, and color superconductors as quark matter \cite{Balachandran}. 
It has been predicted that the relation [Eq. \eqref{eq:single-component-universal-relation}] is changed for superfluid systems with internal degrees of freedom inducing vortices having fractional circulations \cite{Stein,Korshunov}.
However the existence of an unusual BKT transition is not yet conclusive.

Here we consider a two-dimensional Bose system with two different quantum sublevels.
We consider the situation in which two phases in the both sublevels can be synchronized through the Josephson coupling.
When the Josephson coupling is switched off, vortices for both components have fractional circulation $\pm 2 \pi \alpha_i \hbar / m$ ($i = 1, 2$) with the fractional parameter $\alpha_i \in (0,1)$ for the $i$th component with $\alpha_1 + \alpha_2 = 1$.
Under a finite Josephson coupling, which makes superfluid currents of both components the same spatial profiles, a single vortex in each component cannot exist.
Instead, two vortices in both components are connected \cite{Son,Tanaka:2001} to form a vortex molecule as a stable state having the circulation $\pm 2 \pi \hbar / m$ \cite{Son,Kasamatsu-3}.
Dynamics of such vortex molecules have been studied in Refs.~\cite{Tylutki:2016mgy,Eto:2017rfr} 
and vortex lattices have been studied in Refs.~\cite{Cipriani,Uranga}.

The similar situation can appear in a spin-1 spinor Bose system under the quadratic Zeeman effect \cite{James,Kobayashi}, where topologically unstable half-quantized vortices and topologically stable integer vortices appeared (see Appendix \ref{sec:2comp-vs-polar}).
As well as the Josephson coupling in the two-component Bose system, the quadratic Zeeman effect connects two half-quantized vortices with a kink to a integer vortex.
For this system, the two-step phase transitions has been reported \cite{James} due to unbindings of half-quantized vortices and integer vortices with the unconventional jump of the superfluid density.

In this Letter, 
we investigate a possibility of the BKT transition in this system and obtain the following results.
When the Josephson coupling is switched off, there are two-step BKT transitions induced by bindings of fractional vortex-antivortex pairs of each component. % regardless of the inter-component interaction.
Both the BKT transition temperatures depend on the fractional parameter $\alpha_i$.
The jump of the superfluid density becomes twice of the right-hand side in Eq. \eqref{eq:single-component-universal-relation} only when $\alpha_1 = \alpha_2 = 1/2$. %, and remains unchanged otherwise.
On the other hand, the jump of the superfluid density is unchanged when $\alpha_1 \neq \alpha_2$, which suggests that the fractional circulation is never sufficient condition for the change of the jump of the superfluid density [Eq. \eqref{eq:single-component-universal-relation}] opposed to predictions for superfluid $^3$He \cite{Stein,Korshunov} and the spinor Bose system \cite{Mukerjee,James}.
When the Josephson coupling is switched on, the BKT transition is induced by bindings of molecule-antimolecule pairs with the normal jump of the superfluid density [Eq. \eqref{eq:single-component-universal-relation}], while the bindings of vortex-antivortex pairs for each component do not give a phase transition, but they do two crossovers in contrast to the case of the spinor Bose system \cite{James}.

Our results can be tested by
%two weakly connected ultracold Bose gases or
an ultracold Bose mixture with two magnetic hyperfine spin sublevels.
For these systems, the intercomponent Josephson coupling can be realized by
%the energy barrier separating two slices or
the Rabi coupling between two sublevels.
The superfluidity of the system and the BKT transition can be experimentally observed by the same technique as the scalar Bose system \cite{Hadzibabic}.
Because different components do not create the interference pattern in principle, only the global phase of two components contributing to the superfluid density can be observed.
This system has also been considered as a toy system simulating the quark confinement in quantum chromodynamics (QCD) 
\cite{Tylutki:2016mgy,Eto:2017rfr}, thereby our result could give some implications for statistical properties such as a transition between confinement and deconfinement phases.
Other candidates are multiband superconductors such as superconducting MgB$_2$ compounds \cite{Liu,Brinkman,Golubov,Choi} and iron-based superconductors \cite{Kamihara,Singh,Mazin}; it is predicted that the first one has less Josephson coupling strength than the second one.
Although the above materials have not been completely confirmed as multiband superconductors yet, 
our results would give some guiding principles.

%This paper is organized as follows.
%In Sec. \ref{sec:model}, we give the model of our work and discuss vortices, interactions between them, the symmetry of the Hamiltonian and the quasi long-range order appearing below the BKT transition temperature.
%We show our numerical results in Sec. \ref{sec:numerical-result}.
%Section \ref{sec:summary} is devoted to a summary and discussion.

%\section{Model} \label{sec:model}

We start from the Hamiltonian $H = \int d^2x\: \mathcal{H}$ as
\begin{align}
\begin{split}
\mathcal{H} &= \sum_{i = 1}^2 \left\{ \frac{\hbar^2}{2 m} |\nabla \psi_i|^2 + \frac{g_1}{2} |\psi_i|^4 \right\} \\
&\quad + g_{2} |\psi_1|^2 |\psi_2|^2 - \frac{q}{2} \left( \psi_1^\ast \psi_2 + \psi_2^\ast \psi_1 \right),
\end{split}
\label{eq:Hamiltonian}
\end{align}
for two-dimensional Bose mixtures describing two different quantum sublevels coupled by the Josephson coupling.
Here $\psi_i$ ($i = 1,2$) is the $i$th Bose field with the particle mass $m$, $g_1 > 0$ is the intracomponent interaction strength common for both components,
$g_2 > 0$ is the intercomponent interaction strength, and $q \geq 0$ is the Josephson coupling strength.
Here, we set $g_1 > g_2$ for the miscible ground state.
Considering a BKT transition as a phenomenon at finite temperatures, we ignore the quantum fluctuation.
%The thermal average $\langle f \rangle$ of a physical observable $f[\psi_i, \psi_i^\ast]$ at the temperature $T$ is defined as
%\begin{align}
%\langle f \rangle \equiv
%\frac{\displaystyle \int D\psi_1 D\psi_1^\ast D\psi_2 D\psi_2^\ast\: f e^{- \mathcal{H} / T}}
%{\displaystyle \int D\psi_1 D\psi_1^\ast D\psi_2 D\psi_2^\ast\: e^{- \mathcal{H} / T}},
%\label{eq:thermal-average}
%\end{align}
We further impose an additional constraint $(1 / L^2) \int d^2x\: |\psi_i|^2 = n_i$ ($i = 1,2$), where $n_i$ and $L$ are the particle number density for the $i$th component and the system size, respectively.

Inserting the uniform ground state $\psi_i = \sqrt{n_i} e^{i \varphi_i}$ with the phase $\varphi_i$ ($i = 1,2$) for the $i$th component into the Hamiltonian [Eq. \eqref{eq:Hamiltonian}], we obtain the energy density
\begin{align}
\mathcal{E} &= \frac{g_1 n^2}{2} - (g_{1} -g_{2}) \tilde{n}^2 - q \tilde{n} \cos\Delta\varphi,
\label{eq:energy-density}
\end{align}
where $n = n_1 + n_2$, $\tilde{n} = \sqrt{n_1 n_2}$, and $\Delta\varphi = \varphi_1 - \varphi_2$ are the total number density, the geometric mean density, and the relative phase, respectively.
The last term of the right-hand side in Eq.~\eqref{eq:energy-density} shows that the zero relative phase $\varphi_1 - \varphi_2 = 0$ is selected for the ground state.

%\subsection{Vortex, vortex pair, vortex molecule, and molecule pair} \label{subsec:vortex-species}

We first consider vortices and interactions between them.
For $q = 0$, single vortex states
\begin{align}
\{\pm 1,0\} : \left( \psi_1 \sim \sqrt{n_1}\: e^{\pm i \theta_0},\ \psi_2 \sim \sqrt{n_2} \right),
\label{eq:vortex-state-1}
\end{align}
and
\begin{align}
\{0,\pm 1\} : \left( \psi_1 \sim \sqrt{n_1},\ \psi_2 \sim \sqrt{n_2}\: e^{\pm i \theta_0} \right),
\label{eq:vortex-state-2}
\end{align}
are topologically stable, with $\theta_0 \equiv \tan^{-1}(y / x)$.
The mass circulation $\kappa$ of vortices is given by
\begin{align}
\begin{split}
\kappa
%&= \oint d\Vec{l} \cdot \Vec{v} \\
&= \frac{\hbar}{m} \oint d\Vec{l} \cdot \left( \frac{|\psi_1|^2 \nabla \varphi_1 + |\psi_2|^2 \nabla \varphi_2}{|\psi_1|^2 + |\psi_2|^2} \right),
\end{split}
\end{align}
%where $\Vec{v} = (\hbar / m) ( |\psi_1|^2 \nabla \varphi_1 + |\psi_2|^2 \nabla \varphi_2) / (|\psi_1|^2 + |\psi_2|^2)$ is the superfluid velocity, and $\Vec{l}$ is the vector for the closed path surrounding the vortex. 
where $\Vec{l}$ is the vector for the closed path surrounding a vortex. 
The circulation is $\kappa_1 = \pm 2 \pi \alpha_1 \hbar / m$ for vortices $\{\pm 1, 0\}$ and $\kappa_2 = \pm 2 \pi \alpha_2 \hbar / m$ for vortices $\{0,\pm 1\}$, where $\alpha_i \equiv n_i / n$ ($i = 1,2$) is the fractional parameter for the $i$th component.
%For the case of balanced intra-component interaction strengths $\theta = 45^\circ$, circulations of vortices $[\pm 1, 0]$ and $[0, \pm 1]$ are $\kappa = \pm \pi \hbar / m$ with $n_1 = n_2$, where the value $\pi \hbar / m$ is the half of the circulation for the single-component Bose system.
%Vortices as shown in Eqs. \eqref{eq:vortex-state-1} and \eqref{eq:vortex-state-2} are called half-quantized vortices.
%For the case of imbalanced intra-component interaction strengths $\theta < 45^\circ$, circulations of vortices $[0, \pm 1]$ are $\kappa = \pm 2 \pi \hbar / m$ with $n_1 = 0$ and $n_2 \neq 0$ at temperatures between two BKT critical temperatures.
The interaction between the 
$\{\pm 1, 0\}$ and $\{0, \pm 1\}$ vortices are weaker than logarithm 
\cite{Eto:2011wp}, 
and real time dynamics of them have been studied in Ref.~\cite{Kasamatsu:2015cia}.

\begin{figure}[tbh]
\centering
\vspace{\baselineskip}
\begin{minipage}{0.49\linewidth}
\centering
\includegraphics[width=0.99\linewidth]{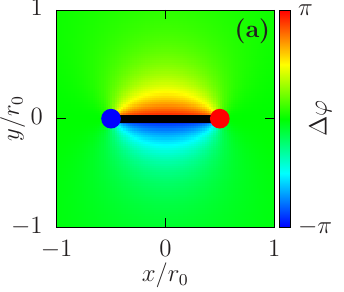}
\end{minipage}
\begin{minipage}{0.49\linewidth}
\centering
\includegraphics[width=0.99\linewidth]{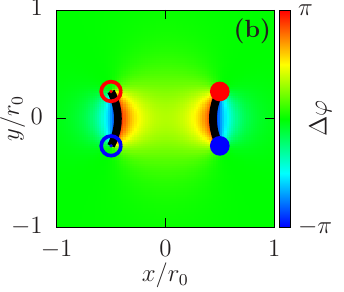}
\end{minipage}
\caption{
\label{fig:molecule}
Profile of the relative phase $\Delta\varphi$ for (a) the vortex molecule $[1,1]_{r_0}$ and the vortex-molecule pair $[1,1]_\delta\:\oset{$r_0$}{-}\:[-1,-1]_\delta$ with $\delta = 0.5 r_0$.
Closed (open) red and blue circles represent positions of vortices of $\{1,0\}$ and $\{0,1\}$ ($\{-1,0\}$ and $\{0,-1\}$), respectively.
Black solid lines show kinks having the relative phase $\Delta\varphi = \pi$.
Both panels are obtained by numerically minimizing the Hamiltonian [Eq. \eqref{eq:Hamiltonian}] with fixing positions of vortices
}
\end{figure}

For $q > 0$, vortices $\{\pm 1, 0\}$ and $\{0, \pm 1\}$ are no more topologically stable.
% because the energy density depends on the direction.
A stable topological defect with nonzero Josephson coupling strength $q > 0$ is a vortex molecule $[1,1]_{r_0}: \{1,0\} \:\oset{$r_0$}{-}\: \{0,1\}$ or its antimolecule $[-1,-1]_{r_0}: \{-1,0\} \:\oset{$r_0$}{-}\: \{0,-1\}$, where $A\:\oset{$r_0$}{-}\:B$ indicates that two vortices $A$ and $B$ are placed with the distance $r_0$.
The circulation of the vortex molecule $[1,1]_{r_0}$ is $\kappa_{\rm M} = 2 \pi \hbar / m$, which is the same as that for a single-component Bose system.
The profile of the relative phase $\Delta\varphi$ for a vortex molecule is illustrated in Fig. \ref{fig:molecule}.
A kink structure having the relative phase $\Delta\varphi = \pi$ appears and mediate an attractive force between two vortices that is balanced with repulsion \cite{Kasamatsu-3}.
There are two characteristic structures for defect-antidefect pairs.
The first is a single component vortex-antivortex pair $\{1,0\}\:\oset{$r_0$}{-}\:\{-1,0\}$ or $\{0,1\}\:\oset{$r_0$}{-}\:\{0,-1\}$.
%\begin{align}
%[0,1] - [0,-1] \equiv \left( \sqrt{n_1} ,\ \sqrt{n_2} e^{i \{ \phi_1(r_0) - \phi_1(-r_0) \}} \right).
%\end{align}
The profile of the relative phase for the single-component vortex-antivortex pair is almost same as that shown in Fig. \ref{fig:molecule} (a), providing the kink between the pair.
The second is a molecule-antimolecule pair $[1,1]_\delta\:\oset{$r_0$}{-}\:[-1,-1]_\delta$ with $\delta < r_0$ which is illustrated in Fig. \ref{fig:molecule} (b).

%\subsection{Interaction between vortices} \label{subsec:vortex-interaction}

The interaction energy $E_{\rm int}$ between vortices is given by inserting an ansatz [Eqs. \eqref{eq:vortex-state-1} and \eqref{eq:vortex-state-2}] into the Hamiltonian [Eq. \eqref{eq:Hamiltonian}].
The leading term in the large $r_0$ limit becomes
\begin{subequations}
\begin{align}
& E_{\rm int}([1,1]_{r_0}) \sim \varepsilon r_0,
\label{eq:interaction-molecule} \\
& E_{\rm int}(\{1,0\}\:\oset{$r_0$}{-}\: \{-1,0\}) \sim \hbar \kappa_1 n \log(r_0 / \xi_1) + \varepsilon r_0,
\label{eq:interaction-pair-1} \\
& E_{\rm int}(\{0,1\}\:\oset{$r_0$}{-}\:\{0,-1\}) \sim \hbar \kappa_2 n \log(r_0 / \xi_2) + \varepsilon r_0,
\label{eq:interaction-pair-2} \\
& E_{\rm int}([1,1]_\delta\:\oset{$r_0$}{-}\:[-1,-1]_\delta) \sim \hbar \kappa_{\rm M} n \log(r_0 / \xi),
\label{eq:interaction-molecule-pair}
\end{align}
\label{eq:vortex-interactions}
\end{subequations}
where $\varepsilon = \gamma \hbar \sqrt{q n \tilde{n} / m}$ is the energy density of the kink per unit length with $\gamma = O(1)$, $\xi_{i}$ ($i = 1,2$) is the vortex core size for the $i$th component, and $\xi = \sqrt{\xi_1 \xi_2}$.
For $g_{1} \gg g_{2}$, $\xi_i$ is estimated as $\xi_i \sim \hbar / \sqrt{2 m g_1 n_i}$.
In order to make the system to exhibit the BKT transition, a pure logarithmic interaction between defect-antidefect pairs is needed.
For $q = 0$, single component vortex-antivortex pairs can induce the BKT transition.
The BKT transition temperatures depend on the vortex core sizes $\xi_i$ and we expect the two-step BKT transitions for the imbalanced density to be $n_1 \neq n_2$.
%The BKT transition temperature increases with decreasing the vortex core size, and we expect that the BKT transition for the second component occurs at higher temperature that for the BKT transition for the first component when $n_1 < n_2$.
For $q > 0$, additional linear terms in Eqs. \eqref{eq:interaction-pair-1} and \eqref{eq:interaction-pair-2} hinder the BKT transitions by bindings of single component vortex-antivortex pairs connected with kinks.
Instead, we expect a new BKT transition by bindings of molecule-antimolecule pairs due to the logarithmic interaction between them in Eq.~\eqref{eq:interaction-molecule-pair}.

We here show our numerical results by using the standard Monte Carlo sampling for the superfluid number density $\rho_{\rm s}$ defined as \cite{Thijssen}
\begin{align}
\rho_{\rm s} = \frac{2 m}{\hbar^2 L^2} \lim_{\Delta \to 0} \frac{F(\Delta) - F(0)}{\Delta^2},
\end{align}
where $F(\Delta)$ is the free energy $- T \log Z(\Delta)$ with the partition function $Z(\Delta) = \langle e^{- H / T} \rangle$ under the twisted boundary condition along the $x$ direction:
$\psi_i(x+L,y) = \psi_i(x,y) e^{i L \Delta}$ \cite{spin-superfluidity}.
For numerical parameters, we use $g_2 = 0.5 g_1$ and $\Delta = 0.01 / \xi$.
\begin{figure}[tbh]
\centering
\vspace{\baselineskip}
\begin{minipage}{0.49\linewidth}
\centering
\includegraphics[width=0.99\linewidth]{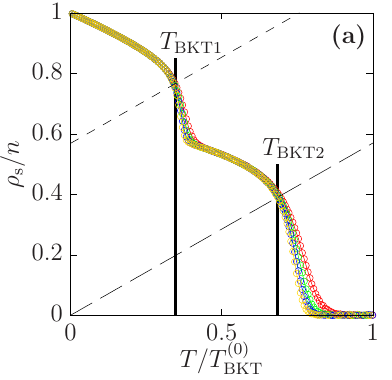}
\end{minipage}
\begin{minipage}{0.49\linewidth}
\centering
\includegraphics[width=0.99\linewidth]{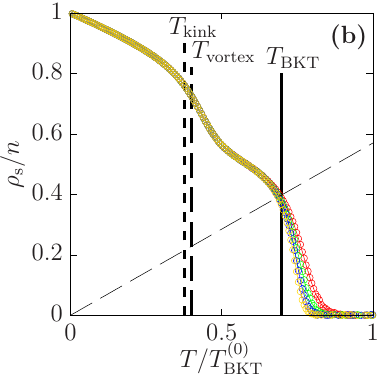}
\end{minipage} \vspace{0.25\baselineskip} \\
\begin{minipage}{0.49\linewidth}
\centering
\includegraphics[width=0.99\linewidth]{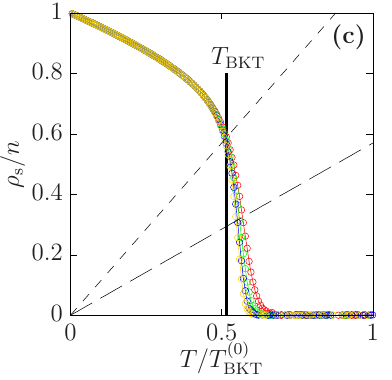}
\end{minipage}
\begin{minipage}{0.49\linewidth}
\centering
\includegraphics[width=0.99\linewidth]{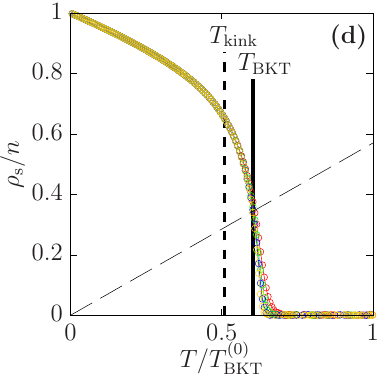}
\end{minipage}
\caption{
\label{fig:rhos}
Superfluid number density $\rho_{\rm s}$ for (a): $n_1 / n_2 = 0.5$ and $q = 0$, (b): $n_1 / n_2 = 0.5$ and $q = 0.1 g_1 n$, (c): $n_1 / n_2 = 1$ and $q = 0$, and (d): $n_1 / n_2 = 1$ and $q = 0.1 g_1 n$ as a function of the temperature $T / T_{\rm BKT}^{(0)}$, where $T_{\rm BKT}^{(0)}$ is the BKT transition temperature for the single-component Bose system..
The system sizes are $L = 64 \xi$ (red lines), $L = 96 \xi$ (green lines), $L = 128 \xi$ (blue lines), and $L = 160 \xi$ (yellow lines). 
The BKT transition temperatures $T_{\rm BKT1}$, $T_{\rm BKT2}$ and $T_{\rm BKT}$ are shown as the thick solid lines.
The thin dashed lines in the all panels show the relation $\rho_{\rm s} / T = 2 m / (\pi \hbar^2)$.
The thin dotted lines in panels (a) and (c) show relations $(\rho_{\rm s} - 0.57 n)/ T = 2 m / (\pi \hbar^2)$ and $\rho_{\rm s} / T = 4 m / (\pi \hbar^2)$, respectively.
The thick dashed line in panel (b) and dotted lines in panels (b) and (d) show the crossover temperatures $T_{\rm vortex}$ and $T_{\rm kink}$, respectively (see Fig. \ref{fig:defect}).
}
\end{figure}
Figure \ref{fig:rhos} shows the temperature dependences of the superfluid density $\rho_{\rm s}$ with various system sizes $L$.
In Fig.~\ref{fig:rhos} (a), there is a strong system size dependence of the superfluid density $\rho_{\rm s}$ just above two BKT transition temperatures $T_{\rm BKT1}$ and $T_{\rm BKT2}$, 
which can be estimated from the binder ratio (see Appendix \ref{sec:binder-ratio}).
In the thermodynamic limit $L \to \infty$, the behavior of the superfluid density $\rho_{\rm s}$ converges to a discrete jump at the BKT transition temperatures.
In Fig.~\ref{fig:rhos} (b), for a nonzero Josephson coupling strength, the system size dependence of the superfluid density at around $T_{\rm BKT} \sim 0.35 T_{\rm BKT}^{(0)}$ disappears, where $T_{\rm BKT}^{(0)}$ is the BKT transition temperature for the single-component Bose system. %, suggesting that the disappearance of the BKT transition at low temperatures.
The rapidly decreasing structure of the superfluid number density still remains above the temperature $T_{\rm vortex} \sim 0.30 T_{\rm BKT}^{(0)}$.
The meaning of the temperature $T_{\rm vortex}$ is related to bindings of vortex-antivortex pairs for the first component, 
as explained later.
For the balanced density $n_1 / n_2 = 1$, there is only one jump structure of the superfluid density for both zero [Fig. \ref{fig:rhos} (c)] and nonzero [Fig. \ref{fig:rhos} (d)] Josephson coupling strengths.

%\subsection{Universal relation at the BKT transition temperature}

%For the single component Bose system, there is a universal relation \eqref{eq:single-component-universal-relation} between the BKT transition temperature $T_{\rm BKT}$ and the jump $\Delta \rho_{\rm s}$ of the superfluid density at the BKT transition temperature $T_{\rm BKT}$.
The universal relation in Eq.~\eqref{eq:single-component-universal-relation} can be confirmed in Fig.~\ref{fig:rhos}: 
the superfluid density $\rho_{\rm s}$ and dashed lines for $\rho_{\rm s} / T = 2 m / (\pi \hbar^2)$ intersect at the BKT transition temperature $T = T_{\rm BKT2}$ in Fig. \ref{fig:rhos} (a) and $T_{\rm BKT}$ in Figs. \ref{fig:rhos} (b) and \ref{fig:rhos} (d).
For $T_{\rm BKT1}$ in Fig. \ref{fig:rhos} (a), we have the same universal relation, i.e., $\rho_{\rm s}$ and the dotted line for $(\rho_{\rm s} - 0.57 n) / T = 2 m / (\pi \hbar^2)$ intersect at $T = T_{\rm BKT1}$, where the density $0.57 n$ is the estimated superfluid density $\rho_{\rm s}$ at the temperature just above $T_{\rm BKT1}$ in the thermodynamic limit $L \to \infty$.
In Fig. \ref{fig:rhos} (c), we have a different universal relation, i.e., $\rho_{\rm s}$ and dotted line for $\rho_{\rm s} / T = 4 m / (\pi \hbar^2)$ intersect at $T_{\rm BKT}$, suggesting two times the right-hand side in Eq.~\eqref{eq:single-component-universal-relation}.
The result shown in Fig. \ref{fig:rhos} (a) suggests that the fractional circulation itself does not affect the universal relation, and the change of the universal relation in Fig. \ref{fig:rhos} (c) can be simply understood by considering that the total jump $\Delta \rho_{\rm s}$ of the superfluid density is separated into two contributions from the both components.
Figures \ref{fig:rhos} (b) and (d) show that defects inducing the BKT transition change from fractional vortices of each component to vortex molecules and molecule-antimolecule pairs, 
also supporting the universal relation [Eq. \eqref{eq:single-component-universal-relation}].

%\subsection{Density of vortices and kinks} \label{subsec:vortex}

\begin{figure}[tbh]
\centering
\vspace{\baselineskip}
\begin{minipage}{0.49\linewidth}
\centering
\includegraphics[width=0.99\linewidth]{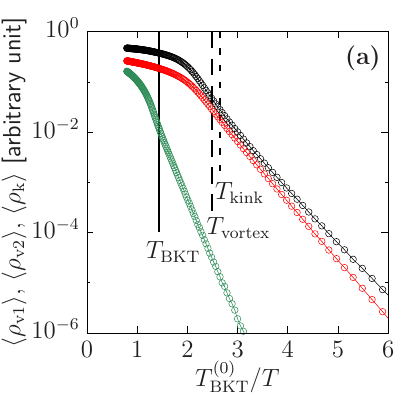}
\end{minipage}
\begin{minipage}{0.49\linewidth}
\includegraphics[width=0.99\linewidth]{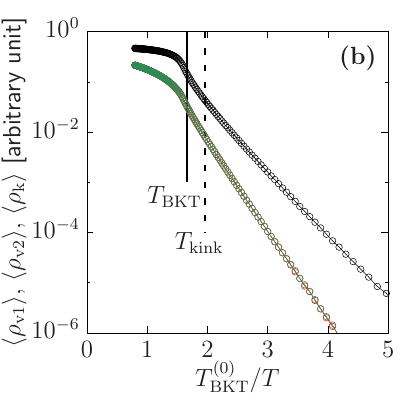}
\end{minipage}
\caption{
\label{fig:defect}
Arrhenius plots for number densities $\langle \rho_{\rm v1} \rangle$ and $\langle \rho_{\rm v2} \rangle$ of vortex-antivortex pairs for first (red lines) and second (green lines) components, and length density $\langle \rho_{\rm k} \rangle$ of kinks (black lines)
for (a): $n_1 / n_2 = 0.5$ and $q = 0.1 g_1 n$ and (b): $n_1 / n_2 = 1$ and $q = 0.1 g_1 n$.
The BKT transition temperature $T_{\rm BKT}$ are shown as solid lines.
  The dashed line in panel (a) and dotted lines in both panels show the crossover temperatures $T_{\rm vortex}$ and $T_{\rm kink}$ above which $\langle \rho_{\rm v1} \rangle$ and $\langle \rho_{\rm k} \rangle$ have deviations larger than 1\% from the Arrhenius relation, respectively.
}
\end{figure}
We next calculate the thermal average of the number density $\langle \rho_{\rm v1} \rangle$ ($\langle \rho_{\rm v2} \rangle$) of vortex-antivortex pairs for the first (second) component and the length density $\langle \rho_{\rm k} \rangle$ of kinks.
At low temperatures, three densities satisfy the Arrhenius relation $\langle \rho_{\rm v1, v2, k} \rangle \propto e^{- \varepsilon_{\rm v1, v2, k} / T}$.
At high temperatures, they
deviate from the Arrhenius relation. % because vortex-antivortex pairs start to unbind and interactions between them are not negligible.
Figure \ref{fig:defect} shows the Arrhenius plots for densities $\rho_{\rm v1, v2, k}$.
With the imbalanced density $n_1 / n_2 = 0.5$ shown in Fig.~\ref{fig:defect} (a), the number density $\langle \rho_{\rm v2} \rangle$ (green line) deviates from the Arrhenius relation at the BKT transition temperature $T_{\rm BKT}$ due to the unbinding of the vortex-antivortex pairs for the second component.
The number density $\rho_{\rm v1}$ (red line) and the length density $\rho_{\rm k}$ (black line), on the other hand, deviate from the Arrhenius relation at lower temperatures $T \simeq 0.40 T_{\rm BKT}^{(0)} = 0.58 T_{\rm BKT} \equiv T_{\rm vortex}$ for $\rho_{\rm v1}$ and $T \simeq 0.38 T_{\rm BKT}^{(0)} = 0.54 T_{\rm BKT} \equiv T_{\rm kink}$ for $\rho_{\rm k}$.
%At $T_{\rm vortex}$, unbinding of vortex-antivortex pairs for the first component does not induce the BKT transition because of the additional linear interaction in Eq. \eqref{eq:interaction-pair-1} but the rapid decrease of the superfluid density $\rho_{\rm s}$ (see Fig. \ref{fig:rhos} (b)).
The temperature $T_{\rm vortex}$ does not induce the BKT transition because of the additional linear interaction in 
Eq.~\eqref{eq:interaction-pair-1}, but shows the crossover for unbinding of vortex-antivortex pairs for the first component with the rapid decrease of the superfluid density $\rho_{\rm s}$ [see Fig. \ref{fig:rhos} (b)].
At the BKT transition temperature $T_{\rm BKT}$, some unbounded vortex-antivortex pairs for the first component couple to those for the second component and form molecule-antimolecule pairs.
The temperature $T_{\rm kink}$ also shows the crossover for nucleation of kink rings with no attached vortices, and $T_{\rm kink} = 0$ with the zero Josephson coupling strength $q = 0$ because there is no energy cost for kinks to be nucleated. 
We note that overall behaviors shown in Figs. \ref{fig:rhos} (a), \ref{fig:rhos} (b), and \ref{fig:defect} (c) are qualitatively unchanged among different imbalanced densities $n_1 / n_2 \neq 1$. 

\begin{figure}[tbh]
\centering
\vspace{\baselineskip}
\begin{minipage}{0.49\linewidth}
\centering
\includegraphics[width=0.99\linewidth]{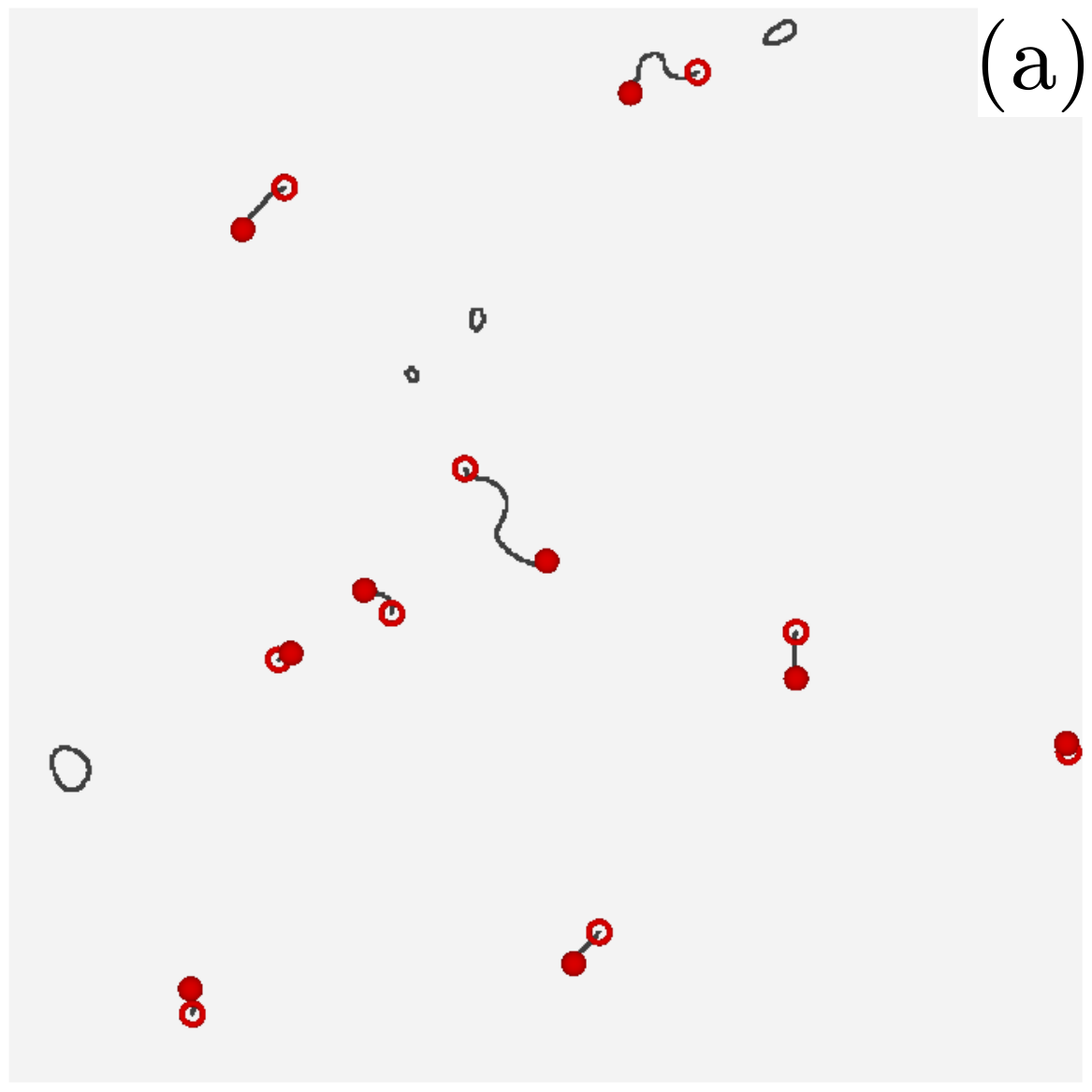}
\end{minipage}
\begin{minipage}{0.49\linewidth}
\includegraphics[width=0.99\linewidth]{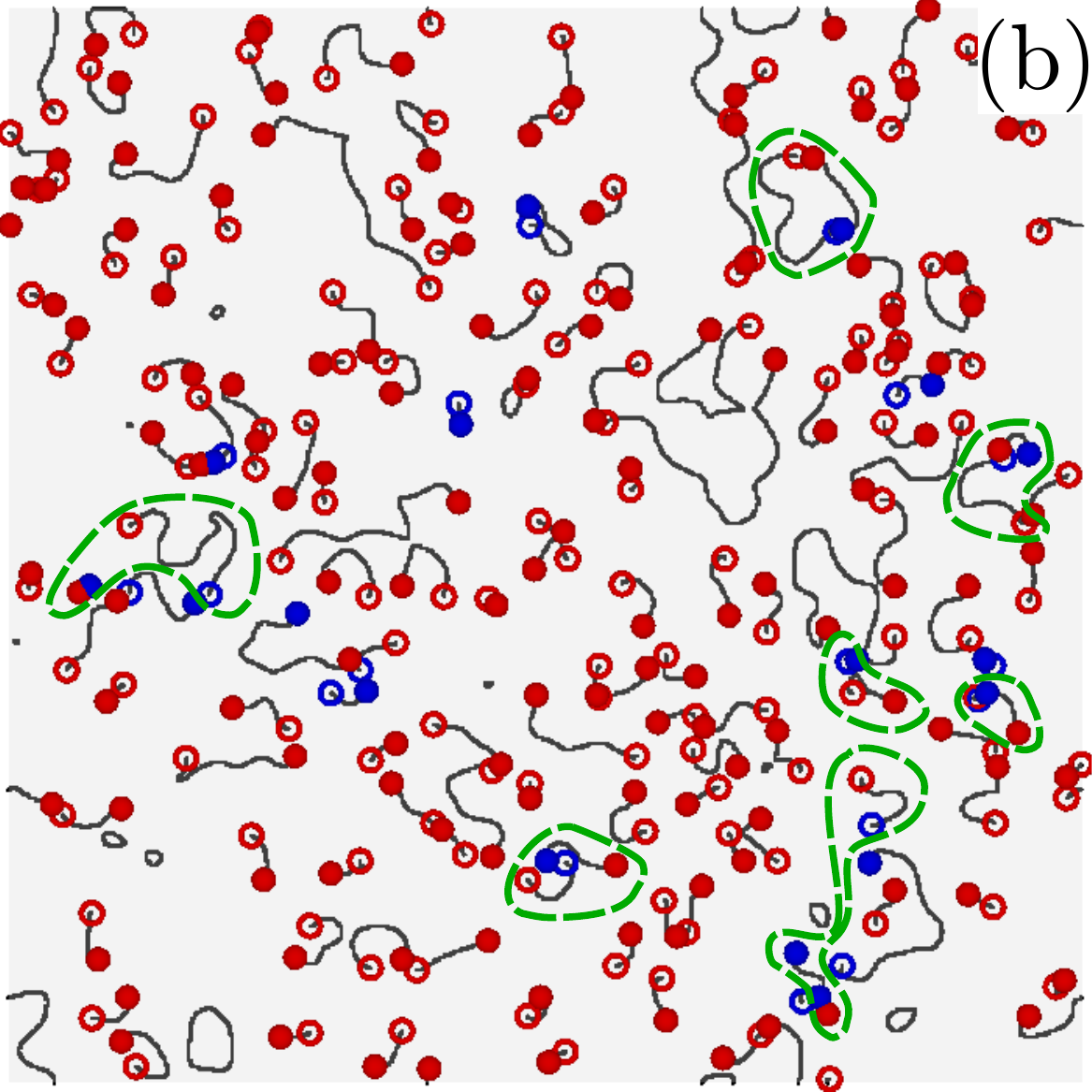}
\end{minipage}
\caption{
\label{fig:vortex}
Equilibrium snapshots of vortices and kinks with $L = 64 \xi$, $n_1 / n_2 = 0.5$ and $q = 0.1 g_1 n$ at
%(a): $T = 0.38 T_{\rm BKT}^{(0)} = \tilde{T}$ and (b): $T = 0.70 T_{\rm BKT}^{(0)} = T_{\rm BKT}$.
(a): $T = T_{\rm vortex}$ and (b): $T = T_{\rm BKT}$.
Closed (open) red and blue circles represent positions of vortices for $\{1,0\}$ and $\{0,1\}$ ($\{-1,0\}$ and $\{0,-1\}$) respectively.
Black solid lines show kinks.
Green dashed closed lines in panel (b) denote molecule-antimolecule pairs.
}
\end{figure}
Figure \ref{fig:vortex} shows equilibrium snapshots of vortices and kinks.
At $T = T_{\rm vortex}$, there are vortex-antivortex pairs for the first component connected with kinks as shown in \ref{fig:vortex} (a).
At $T = T_{\rm BKT}$, there are not only vortex-antivortex pairs for both components but also molecule-antimolecule pairs denoted by green dashed closed lines.
In both Figs. \ref{fig:vortex} (a) and \ref{fig:vortex} (b), we can see kink rings with no attaching vortices which start to frequently appear at $T_{\rm kink}$.

For the balanced density $n_1 = n_2$ shown in Fig.~\ref{fig:defect} (b), two number densities $\rho_{\rm v1} \simeq \rho_{\rm v2}$ and the length density $\rho_{\rm k}$ deviate from the Arrhenius relation at the BKT transition temperature $T_{\rm BKT}$ ($= T_{\rm vortex}$) and the crossover temperature $T_{\rm kink}$, respectively.

\begin{figure}[tbh]
\centering
\vspace{\baselineskip}
\begin{minipage}{0.49\linewidth}
\centering
\includegraphics[width=0.99\linewidth]{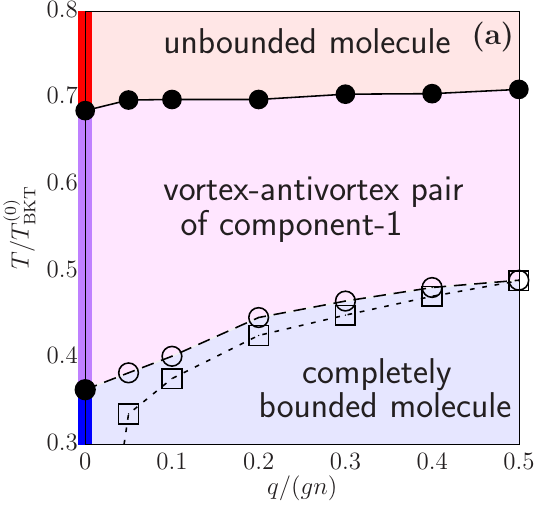}
\end{minipage}
\begin{minipage}{0.49\linewidth}
\includegraphics[width=0.99\linewidth]{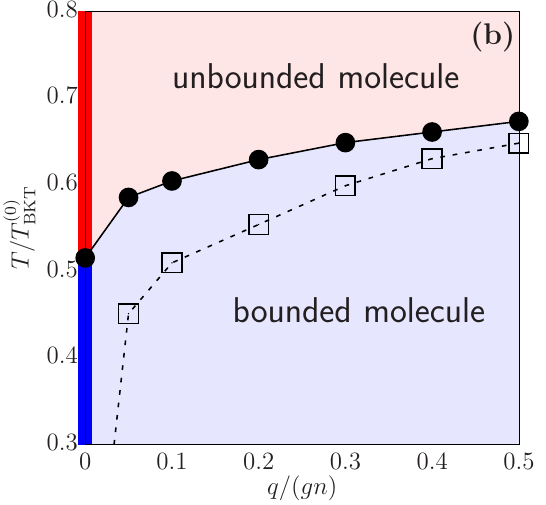}
\end{minipage}
\caption{
\label{fig:phase-diagram}
Phase diagrams for the temperature $T$ and the Josephson coupling $q$ for (a): $n_1 / n_2 = 0.5$ and (b): $n_1 / n_2 = 1$.
The red, purple, and blue lines at $q = 0$ show phases for unbounded vortex of both components, unbounded vortex of the first component, and bounded vortex of both components, respectively.
The black solid lines and filled circles show the BKT transition temperature $T_{\rm BKT}$.
The dashed line with open circles in panel (a) and dotted lines with open squares in the both panels indicate the crossover temperatures $T_{\rm vortex}$ and $T_{\rm kink}$.
}
\end{figure}
%We finally show the phase diagram for the temperature $T$ and the Josephson coupling $q$ in Fig. \ref{fig:phase-diagram}.
%The structure of the phase diagram is completely divided between $q = 0$ and $q > 0$, i.e., the BKT transition is induced by fractional vortices at $q = 0$ and vortex molecules at $q > 0$.
%For the imbalanced density $n_1 < n_2$, three thermodynamic phases separated by BKT transitions at $q = 0$ changes to two thermodynamic phases for bounded and unbounded molecules.
%The former one is further separated into two regions by the temperature $\tilde{T}$ for bounded and unbounded vortices of the first component, the boundary of which.
%For the balanced density $n_1 = n_2$, the phase at $q = 0$ (purple line) and the region at $q > 0$ (purple area) for unbounded vortices of the first component (purple line) shrink and there are two thermodynamic phases for both $q = 0$ and $q > 0$.
%
%%\subsection{Summary and discussion} \label{sec:summary}
%
In conclusion, we have investigated the two-dimensional Bose system with two quantum sublevels.
The phases of both components can be coupled through the Josephson coupling.
When the Josephson coupling is absent, topologically stable fractional vortices induce two-step BKT transitions having the normal universal relation in Eq.~\eqref{eq:single-component-universal-relation}, which suggests that the fractional circulation does not affect the universal relation in contrast to the conventional understanding.
This result is qualitatively independent of the value of the intercomponent interaction strength $g_2$.
%For the imbalanced intra-component interaction strengths, two-step BKT transitions at different temperature give the conventional universal relation with $c = 1$ even though vortices have half-quantized circulation.
%This result suggest that fractional circulations of vortices are not the sufficient condition for the unconventional universal relation with $c \neq 1$.
When the Josephson coupling is switched on, the BKT transition occurs once for both balanced and imbalanced densities, and the universal relation is unchanged even for the balanced density.
This result can be understood from the interaction between vortices as shown in Eq.~\eqref{eq:vortex-interactions}.
There are additional linear interactions between a vortex and its antivortex in Eqs.~\eqref{eq:interaction-pair-1} and \eqref{eq:interaction-pair-2} which hinder the BKT transition by binding of them.
Instead of single-component vortex pairs, molecule-antimolecule pairs having pure logarithmic interactions in Eq.~\eqref{eq:interaction-molecule-pair} induces the BKT transition of this system.
For the case of imbalanced densities, however, there is a characteristic temperature $T_{\rm vortex}$ at which single-component vortex-antivortex pairs start to form bound states as a relic of the BKT transition without the Josephson coupling.
The temperature $T_{\rm vortex}$ does thermodynamically not give the transition but gives the crossover.
We also find a lower crossover temperature $T_{\rm kink}$ than $T_{\rm vortex}$ at which kink rings with no attaching vortices start to be nucleated.
We summarize our discussion with the phase diagrams shown in Fig.~\ref{fig:phase-diagram}.
%
%Our results give some implications to systems having various kinds of topological defects and circulations about possibilities of the BKT transition and their universal relations.
%One important application is

%

%%%%%%%%%%%%%%%%%%%%%%%%%
We would like to thank Yuki Kawaguchi and Atsutaka Maeda for the helpful suggestions and comments. 
This work is supported by 
the Japan Society for the Promotion of Science (JSPS) 
 Grant-in-Aid for Scientific Research (KAKENHI Grant No. 16H03984).
The work is also supported in part by Grant-in-Aid for Scientific Research 
(KAKENHI Grant Numbers 
26870295 (MK), 16KT0127 (MK),
26800119 (ME), 17H06462 (ME),
and 18H01217 (MN)).
The work of MN is supported in part by the Ministry of Education, Culture, Sports, Science (MEXT)-Supported Program for the Strategic Research Foundation at Private Universities ``Topological Science'' (Grant No. S1511006),
and by a Grant-in-Aid for
Scientific Research on Innovative Areas ``Topological Materials
Science'' (KAKENHI Grant No.~15H05855) 
from the the Ministry of Education,
Culture, Sports, Science (MEXT) of Japan.

%%%%%%%%%%%%%%%%%%%%%

%%%%%%%%%%%%%%%%%%%%%


\begin{thebibliography}{100}
\bibitem{spin-superfluidity}
When $q = 0$, we can further define the pseudo-spin superfluid density.
For details, see Appendix \ref{sec:spin-superfluidity}.
\end{thebibliography}

\vspace{2\baselineskip}

%%%%%%%%%%%%%%%%%%%%%%%%%%

\appendix

\section{How to fix the Berezinskii-Kosteritz-Thouless transition temperature}
\label{sec:binder-ratio}

The Binder ratio
\begin{align}
b \equiv 1 - \frac{\langle \{ |\Psi_1|^2 + |\Psi_2|^2 \}^2 \rangle}{3 \langle |\Psi_1|^2 + |\Psi_2|^2 \rangle^2},
\end{align}
becomes size independent for the critical states below the BKT transition temperature.
Here, $\Psi_i$ is defined as
\begin{align}
\Psi_i \equiv \frac{1}{L^2} \int d^2x\: \psi_i.
\end{align}
\begin{figure}[tbh]
\centering
\vspace{\baselineskip}
\begin{minipage}{0.49\linewidth}
\centering
\includegraphics[width=0.99\linewidth]{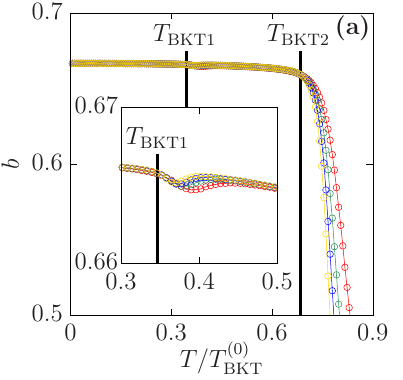}
\end{minipage}
\begin{minipage}{0.49\linewidth}
\centering
\includegraphics[width=0.99\linewidth]{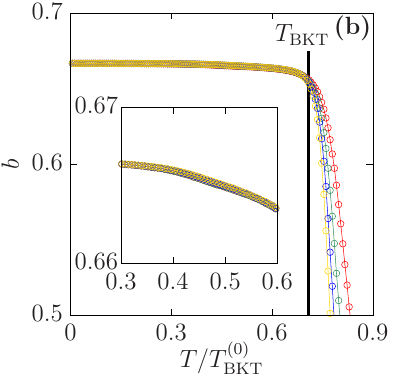}
\end{minipage}
\caption{
\label{fig:binder}
Binder ratio $b$ for $n_1 / n_2 = 0.5$ and system sizes $L_1 = 64 \xi$ (red lines), $L_2 = 96 \xi$ (green lines), $L_3 = 128 \xi$ (blue lines), and $L_4 = 160 \xi$ (yellow lines).
The Josephson coupling strength is fixed as (a) $q = 0$ and (b) $q = 0.1 g_1 n$.
The BKT transition temperatures ($T_{\rm BKT1}$ and $T_{\rm BKT2}$ in panel (a) and $T_{\rm BKT}$ in panel (b)) are shown as solid lines.
In insets for both panels, we zoom over temperatures around $T = 0.4 T_{\rm BKT}^{(0)}$. 
}
\end{figure}
Figure \ref{fig:binder} shows two typical examples of the Binder ratio $b$ with the Josephson coupling strength $q = 0$ and $q = 0.1 g_1 n$ for the imbalanced density $n_1 / n_2 = 0.5$.
With four different system sizes, the BKT transition temperature is estimated as the temperature at which four Binder ratios start to differ:
\begin{align}
\frac{1}{4 \bar{b}} \sqrt{\sum_{i = 1}^4 \left(b(L_i) - \bar{b}\right)^2} \lesssim 0.01, \quad
\bar{b} \equiv \frac{1}{4} \sum_{i = 1}^4 b(L_i).
\end{align}
For the zero Josephson coupling strength $q = 0$ shown in Fig. \ref{fig:binder} (a), there are two transition temperatures $T_{\rm BKT1} \simeq 0.36 T_{\rm BKT}^{(0)}$ and $T_{\rm BKT2} \simeq 0.68 T_{\rm BKT}^{(0)}$ for the first (see the inset in Fig. \ref{fig:binder} (a)) and second components, respectively. %, where $T_{\rm BKT}^{(0)}$ is the BKT transition temperature for the single-component Bose system.
For the nonzero Josephson coupling strength $q = 0.1 g_1 n$ in Fig. \ref{fig:binder} (b), on the other hand, the BKT transition at the low temperature around $T \sim 0.35 T_{\rm BKT}^{(0)}$ disappears and occurs only once at the temperature $T_{\rm BKT} \simeq 0.71 T_{\rm BKT}^{(0)}$.
All other BKT transition temperatures $T_{\rm BKT}$ in the manuscript are obtained by the same way.

Obtained BKT transition temperatures can be also justified by the correlation function
\begin{align}
C(r) = \frac{1}{n L^2} \sum_{i = 1}^2 \int d^2x\: \int \frac{d\Omega_{\Vec{r}}}{2 \pi r}
\left\langle \psi_i^\ast(\Vec{x}) \psi_i(\Vec{x}+\Vec{r}) \right\rangle,
\end{align}
where $\Omega_{\Vec{r}}$ is the solid angle for the vector $\Vec{r}$.
In the thermodynamic limit $L \to \infty$, the correlation function $C(r)$ show the algebraic decrease $C(r) \propto r^{-\eta(T)}$ with the temperature-dependent critical exponent $\eta(T)$ at low temperatures $T \leq T_{\rm BKT}$.
\begin{figure}[tbh]
\centering
\vspace{\baselineskip}
\begin{minipage}{0.49\linewidth}
\centering
\includegraphics[width=0.99\linewidth]{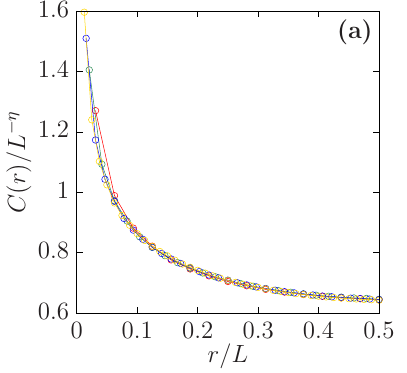}
\end{minipage}
\begin{minipage}{0.49\linewidth}
\centering
\includegraphics[width=0.99\linewidth]{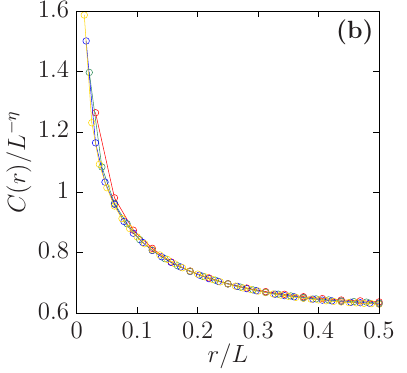}
\end{minipage} \\[10pt]
\begin{minipage}{0.49\linewidth}
\centering
\includegraphics[width=0.99\linewidth]{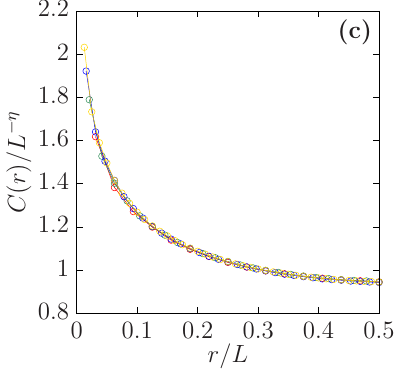}
\end{minipage}
\begin{minipage}{0.49\linewidth}
\centering
\includegraphics[width=0.99\linewidth]{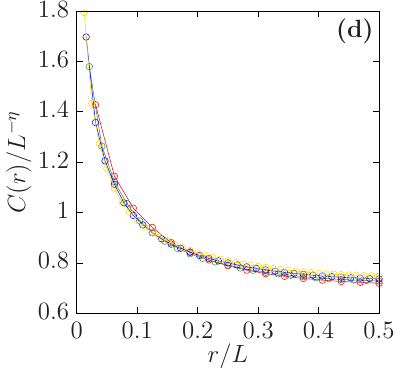}
\end{minipage}
\caption{
\label{fig:correlation-high}
Finite-size scaling of the correlation function $C(r)$ for (a) $n_1 / n_2 = 0.5$ and $q = 0$, (b) $n_1 / n_2 = 0.5$ and $q = 0.1 g_1 n$, (c) $n_1 / n_2 = 1$ and $q = 0$, and (d) $n_1 / n_2 = 1$ and $q = 0.1 g_1 n$ at the BKT transition temperature (a) $T = T_{\rm BKT2}$ and (b-d) $T = T_{\rm BKT}$.
The critical exponent is fixed as $\eta = 1/4$.
For the system size, we use the same colors as those in Fig. \ref{fig:binder}. 
}
\end{figure}
The correlation function $C(r) / L^{-\eta}$ is expected to be a universal function on $r / L$ below BKT critical temperatures.
We can see this universality in Fig. \ref{fig:correlation-high} with $\eta = 1/4$ at $T = T_{\rm BKT2}$ (panel (a)) and $T = T_{\rm BKT}$ (panels (b-d)).
The value of the exponent $\eta = 1/4$ is same as that for the single-component Bose gas at the BKT critical temperature.
Our results in Fig. \ref{fig:correlation-high} support the correctnesses of our estimations for the BKT critical temperatures and suggest that the critical exponent is unchanged between the single-component and multicomponent Bose gases.

Considering the universality at $T = T_{\rm BKT1}$ for the imbalanced density $n_1 / n_2 \neq 1$ and the zero Josephson coupling strength $q = 0$, we should use the correlation function $C_1(r)$ for the 1st component defined as
\begin{align}
C_1(r) = \frac{1}{n_1 L^2} \int d^2x\: \int \frac{d\Omega_{\Vec{r}}}{2 \pi r}
\left\langle \psi_1^\ast(\Vec{x}) \psi_1(\Vec{x}+\Vec{r}) \right\rangle,
\end{align}
instead of the global correlation function $C(r)$.
In the same way as $C_1(r)$, we can define $C_2(r)$ for the second component and its exponent is smaller than $1/4$ because the BKT transition for the second component occurs at the higher temperature $T_{\rm BKT2} > T_{\rm BKT1}$.
\begin{figure}[tbh]
\centering
\vspace{\baselineskip}
\includegraphics[width=0.50\linewidth]{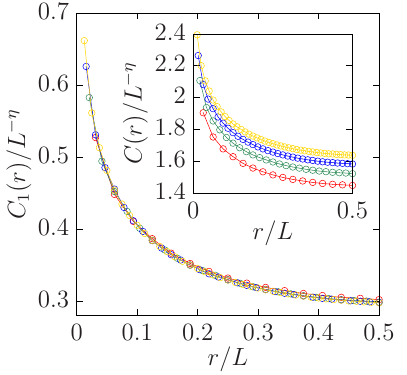}
\caption{
\label{fig:correlation-low}
Finite-size scaling of the correlation function $C_1(r)$ for $n_1 / n_2 = 0.5$ and $q = 0$ at the BKT transition temperature $T = T_{\rm BKT1}$.
The critical exponent is fixed as $\eta = 1/4$.
For the system size, we use the same colors as those in Fig. \ref{fig:binder}.
Inset: Finite-size scaling of the correlation function $C(r)$.
}
\end{figure}
The universality of $C_1(r)$ with $\eta = 1/4$ at $T = T_{\rm BKT}$ can be confirmed as shown in Fig. \ref{fig:correlation-low}, whereas $C(r)$ fails to satisfy the universal property as shown in the inset of Fig. \ref{fig:correlation-low}.

\section{Pseudo-spin superfluid density}
\label{sec:spin-superfluidity}

The (mass) superfluid density in Eq. (8) can be defined by the invariance of the Hamiltonian (1) under a global phase shift $\psi_i \to e^{i \Delta} \psi_i$ ($i = 1,2$).
When the Josephson coupling is switched off as $q = 0$, the Hamiltonian is also invariant under the relative phase shift $\psi_i \to e^{i \sigma_i \Delta} \psi_i$ ($\sigma_1 = 1$ and $\sigma_2 = -1$), which defines the pseudo-spin superfluid density
\begin{align}
\rho_{\rm ps} = \frac{2 m}{\hbar^2 L^2} \lim_{\Delta \to 0} \frac{F_{\rm s}(\Delta) - F(0)}{\Delta^2},
\end{align}
where $F_{\rm s}(\Delta)$ is the free energy under the boundary condition with the relative phase twist $\psi_i(x+L,y) = e^{i L \sigma_i \Delta} \psi(x,y)$.

\begin{figure}[htb]
\centering
\begin{minipage}{0.49\linewidth}
\includegraphics[width=0.95\linewidth]{rhos-0-035.pdf}
\end{minipage}
\begin{minipage}{0.49\linewidth}
\includegraphics[width=0.95\linewidth]{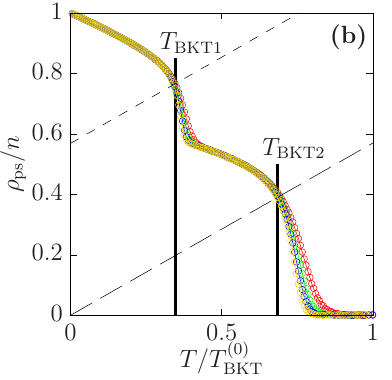}
\end{minipage} \\[10pt]
\begin{minipage}{0.49\linewidth}
\includegraphics[width=0.95\linewidth]{rhos-0-045.pdf}
\end{minipage}
\begin{minipage}{0.49\linewidth}
\includegraphics[width=0.95\linewidth]{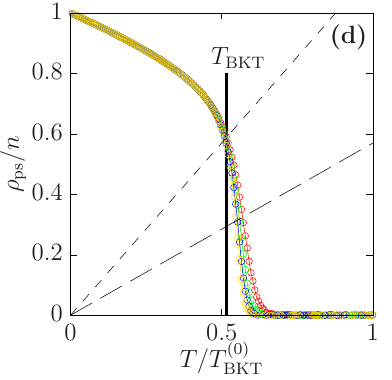}
\end{minipage}
\caption{\label{fig:rhos} Mass superfluid density $\rho_{\rm s}$ (panels (a) and (c)) and pseudo-spin superfluid density $\rho_{\rm ps}$ (panels (b) and (d)) for (a)-(b): $n_1 / n_2 = 0.5$ and (c)-(d) $n_1 /n_2 = 1$ with $q = 0$.}
\end{figure}
Figure \ref{fig:rhos} shows the mass and pseudo-spin superfluid densities $\rho_{\rm s}$ and $\rho_{\rm ps}$.
In both imbalanced ($n_1 \neq n_2$) and balanced ($n_1 = n_2$) cases, the two superfluid densities take the same value within the numerical accuracies.

Our result suggests that properties of superfluidity for both component are independent of each other, and the non-dissipative drag between the two component known as Andreev-Bashkin effect \cite{Nespolo,Karle} is negligible.
The similar result has been reported by a work studying a two-dimensional spin-1 spinor Bose system; the mass and spin superfluid densities take the same value for the antiferromagnetic ground state \cite{Podolsky}.
This work has also found that the difference of two superfluid densities arises when the quantum fluctuation becomes dominant.
Although the quantum fluctuation is usually negligible for phase transitions at finite temperatures, it can be amplified by, for example, the optical lattice potential for the case of ultracold atoms.

\section{Details of numerical procedure}

The thermal average $\langle f \rangle$ of the physical observable $f[\psi_i,\psi_i^\ast]$ defined as
\begin{align}
\langle f \rangle = \frac{\displaystyle \int \left( \prod_{i=1}^2 D \psi_i\: D\psi_i^\ast \right) \left( \int d^2 x\: f \right) e^{- \mathcal{H} / T}}{\displaystyle \int \left(\prod_{i=1}^2 D \psi_i\: D\psi_i^\ast \right) e^{- \mathcal{H} / T}},
\end{align}
can be obtained by the overdamped Langevin equation
\begin{align}
\begin{split}
\frac{d \psi_i}{dt} &= - \frac{\delta \mathcal{H}}{\delta \psi_i} + \sqrt{T} (w_{i,1} + i w_{i,2}) \\
&= \frac{\hbar^2}{2 m} \nabla^2 \psi_i - \left(g_1 |\psi_i|^2 + |\psi_{3-i}|^2 \right) \psi_i + \frac{q}{2} \psi_{3-i} \\
&\quad + \sqrt{T} (w_{i,1} + i w_{i,2}),
\end{split}
\label{eq:Langevin}
\end{align}
where $w_{i,j}$ ($i,j = 1, 2$) is a white Gaussian noise satisfying
\begin{align}
\begin{split}
& \langle w_{i,j}(\Vec{x},t) \rangle = 0, \\
& \langle w_{i,j}(\Vec{x},t) w_{i^\prime,j^\prime}(\Vec{x}^\prime,t^\prime) \rangle = \delta_{i,i^\prime} \delta_{j,j^\prime} \\
& \phantom{\langle w_{i,j}(\Vec{x},t) w_{i^\prime,j^\prime}(\Vec{x}^\prime,t^\prime) \rangle =}
\times \delta(\Vec{x} - \Vec{x}^\prime) \delta(t - t^\prime).
\end{split}
\end{align}
The thermal average $\langle f \rangle$ can be computed as the temporal average of the Langevin equation
\begin{align}
\langle f \rangle = \lim_{\tau \to \infty} \frac{1}{\tau} \int_0^\tau dt\: f.
\end{align}
To numerically solve the Langevin equation [Eq. \eqref{eq:Langevin}], we discretize the space and time as $(\Vec{x}_{m,l},t_n)$.
The Langevin equation becomes
\begin{widetext}
\begin{align}
\begin{split}
\psi_i^{(l,m,n+1)}
&= \psi_i^{(l,m,n)} + \Delta t \left\{ \frac{\hbar^2}{2 m} \nabla^2 \psi_i^{(l,m,n)} - \left( g_1 |\psi_i^{(l,m,n)}|^2 + |\psi_{3-i}^{(l,m,n)}|^2 \right) \psi_i^{(l,m,n)} + \frac{q}{2} \psi_{3-i}^{(l,m,n)} \right\} \\
&\quad + \sqrt{T (\Delta x)^2 \Delta t} \left( R_{i,1}^{(l,m,n)} + i R_{i,2}^{(l,m,n)} \right) \\
&\equiv \psi_i^{(l,m,n)} + \Delta \mathcal{H}_i[\psi_i^{(l,m,n)}] + \sqrt{T (\Delta x)^2 \Delta t} \left( R_{i,1}^{(l,m,n)} + i R_{i,2}^{(l,m,n)} \right),
\label{eq:Langevin-Euler}
\end{split}
\end{align}
\end{widetext}
where, $\psi_i(\Vec{x}_{m,l},t_n) \equiv \psi_i^{(m,l,n)}$, and the white noise $R_{i,j}^{(l,m,n)}$ satisfies
\begin{align}
\begin{split}
& \langle R_{i,j}^{(l,m,n)} \rangle = 0, \\
& \langle R_{i,j}^{(l,m,n)} R_{i^\prime,j^\prime}^{(l^\prime,m^\prime,n^\prime)} \rangle = \delta_{i,i^\prime} \delta_{j,j^\prime} \delta_{l,l^\prime} \delta_{m,m^\prime} \delta_{n,n^\prime}.
\end{split}
\end{align}
The Laplace operator is approximated as
\begin{align}
\begin{split}
&\quad \nabla^2 \psi_i^{(l,m,n)} \\
&= \frac{1}{(\Delta x)^2} \left(\psi_i^{(l+1,m,n)} + \psi_i^{(l-1,m,n)} \right. \\
&\phantom{= \frac{1}{(\Delta x)^2} (} \left. + \psi_i^{(l,m+1,n)} + \psi_i^{(l,m-1,n)} - 4 \psi_i^{(l,m,n)}\right).
\end{split}
\end{align}
We improve the Euler scheme used in Eq. \eqref{eq:Langevin-Euler} for the time derivative as
\begin{align}
\begin{split}
& \Delta \psi_i^{(l,m,n)} = \Delta \mathcal{H}_i[\psi_i^{(l,m,n)}] \\
&\phantom{\Delta \psi_i^{(l,m,n)}} + \sqrt{T (\Delta x)^2 \Delta t} \left\{R_{i,1}^{(l,m,n)} + i R_{i,2}^{(l,m,n)}\right\}, \\
& \tilde{\psi}^{(l,m,n)} = \psi_i^{(l,m,n)} + \Delta \psi_i^{(l,m,n)}, \\
& \Delta \tilde{\psi}_i^{(l,m,n)} = \Delta \mathcal{H}_i[\tilde{\psi}_i^{(l,m,n)}] \\
&\phantom{\Delta \tilde{\psi}_i^{(l,m,n)}} + \sqrt{T (\Delta x)^2 \Delta t} \left\{R_{i,1}^{(l,m,n)} + i R_{i,2}^{(l,m,n)}\right\}, \\
& \psi_i^{(l,m,n+1)}
= \psi_i^{(l,m,n)} + \frac{\Delta \psi_i^{(l,m,n)} + \Delta \tilde{\psi}_i^{(l,m,n)}}{2} \\
&\phantom{\psi_i^{(l,m,n+1)}} \equiv \psi_i^{(l,m,n)} + \Delta \bar{\psi}_i^{(l,m,n)}.
\label{eq:Langevin-improved-Euler}
\end{split}
\end{align}
To impose the constraint
\begin{align}
\frac{(\Delta x)^2}{L^2} \sum_{l,m} |\psi_i^{(l,m,n)}|^2 = n_i,
\end{align}
we introduce the Lagrange multiplier $\mu_i^{(n)}$ as
\begin{align}
\begin{split}
\psi_i^{(l,m,n+1)}
&= \psi_i^{(l,m,n)} + \Delta \bar{\psi}_i^{(l,m,n)} \\
&\quad + \mu_i(t_n) \Delta t \: \bar{\psi}_i^{(l,m,n)} \\
&\equiv F_i^{(l,m,n)} + \mu_i(t_n) \Delta t \: \bar{\psi}_i^{(l,m,n)},
\end{split}
\end{align}
where $\bar{\psi}_i^{(l,m,n)} = \{ \psi_i^{(l,m,n)} + \tilde{\psi}_i^{(l,m,n)} \} /2$.
Requiring $(\Delta x)^2/L^2 \sum_{l,m} |\psi_i^{(l,m,n+1)}|^2 = n_i$, we obtain
\begin{align}
\begin{split}
& \mu_i^{(n)} = \frac{- \mathcal{G}^{(n)}_i + \sqrt{\left(\mathcal{G}^{(n)}_i\right)^2 + \bar{n}^{(n)}_i (n_i - \mathcal{F}^{(n)}_i)}}{\bar{n}^{(n)} \Delta t}, \\
& \mathcal{F}^{(n)}_i = \frac{(\Delta x)^2}{L^2} \sum_{l,m} |F_i^{(l,m,n)}|^2, \\
& \mathcal{G}^{(n)}_i = \frac{(\Delta x)^2}{L^2} \sum_{l,m} \mathrm{Re}[\bar{\psi}_i^{\ast\ (l,m,n)} F_i^{(l,m,n)}], \\
& \bar{n}^{(n)}_i = \frac{(\Delta x)^2}{L^2} \sum_{l,m} |\bar{\psi}_i^{(l,m,n)}|^2.
\end{split}
\end{align}

\section{Dynamics of vortices and kinks}

Dynamics of vortices and kinks obtained by solving the Langevin equation [Eq. \eqref{eq:Langevin-improved-Euler}] can be seen at \\
\url{http://www.ton.scphys.kyoto-u.ac.jp/~michikaz/physics/2-component-BKT.html}

\section{Relation between the two-component Bose system and the polar state of the spin-1 spinor Bose system}
\label{sec:2comp-vs-polar}

The polar states with the negative spin-dependent coupling strength can be written as
\begin{align}
  \psi^{\rm spin-1}(\beta,\gamma,\phi) = e^{i \phi} \sqrt{\frac{n}{2}} \begin{pmatrix} i e^{- i \gamma} \sin\beta \\ \sqrt{2} \cos\beta \\ i e^{i \gamma} \sin\beta \end{pmatrix},
  \label{eq:polar-state}
\end{align}
where $\phi$ is the global phase, and $\beta$ and $\gamma$ are the spin angles.
The Hamiltonian density for the spin-1 spinor Bose system is
\begin{align}
\begin{split}
\mathcal{H}^{\rm spin-1} &= \frac{\hbar^2}{2 m} \left(\nabla \psi^{{\rm spin-1} \dagger}\right) \left(\nabla \psi^{\rm spin-1}\right) \\
&\quad + \frac{1}{2} \left\{ \vphantom{\sum_{\nu = x,y,z}} \lambda_0 \left(\psi^{{\rm spin-1} \dagger} \psi^{\rm spin-1}\right)^2 \right. \\
&\phantom{\quad + \frac{1}{2} \{} \left. + \lambda_1 \sum_{\nu = x,y,z} \left(\psi^{{\rm spin-1} \dagger} s_\nu \psi^{\rm spin-1}\right)^2 \right\} \\
&\quad + q_{\rm Z} \sum_{s=-1}^1 s^2 |\psi^{\rm spin-1}_s|^2,
\end{split}
\label{eq:spinor-Hamiltonian}
\end{align}
where $s_\nu$ ($\nu = x,y,z$) is the spin-1 spin matrix, $\lambda_0$ and $\lambda_1$ are the spin-independent and spin-dependent coupling constant, and $q_{\rm Z}$ is the strength of the quadratic Zeeman effect.
The polar state [Eq. \eqref{eq:polar-state}] is realized when $\lambda_{0,1} > 0$, and we here consider the case $q_{\rm J} > 0$ for the positive quadratic Zeeman effect.
We compare this system with the two-component Bose system with the state
\begin{align}
  \psi^{\rm 2-comp}(\varphi,\Delta\varphi) = e^{i \varphi/2} \begin{pmatrix} \sqrt{n_1} e^{i \Delta\varphi/2} \\ \sqrt{n_2} e^{-i\Delta\varphi/2} \end{pmatrix},
\end{align}
and the Hamiltonian density
\begin{align}
\begin{split}
& \mathcal{H}^{\rm 2-comp} \\
&= \sum_{i=1}^2 \left( \frac{\hbar^2}{2 m} |\nabla \psi^{\rm 2-comp}_i|^2 + \frac{g_1}{2} |\psi^{\rm 2comp}_i|^4 \right) \\
&\quad + g_2 |\psi^{\rm 2-comp}_1|^2 |\psi^{\rm 2-comp}_2|^2 \\
&\quad - \frac{q_{\rm J}}{2} \psi^{{\rm 2-comp} \dagger} \sigma_x \psi^{\rm 2-comp},
\end{split}
\label{eq:2comp-Hamiltonian}
\end{align}
where $\varphi$ and $\Delta \varphi$ are the global and relative phases between the components.
Assuming that $n$ and $n_{1,2}$ are constant, we write the Hamiltonian densities $\mathcal{H}^{\rm spin-1}$ and $\mathcal{H}^{\rm 2-comp}$ with the phase and spin angles as
\begin{align}
\begin{split}
& \mathcal{H}^{\rm spin-1} = \frac{\hbar^2 n}{2 m} \left\{ \left(\nabla \phi\right)^2 + \left(\nabla \beta\right)^2 + \sin^2\beta \left(\nabla \gamma\right)^2 \right\} \\
&\phantom{\mathcal{H}^{\rm spin-1} =} + \frac{\lambda_0 n^2}{2} + \frac{q_{\rm Z} n \{1 - \cos(2 \beta)\}}{2}, \\
& \mathcal{H}^{\rm 2-comp} = \frac{\hbar^2 n}{8 m} \left[ (n_1 + n_2) \left\{ \left(\nabla \varphi\right)^2 + \left(\nabla \Delta\varphi\right)^2 \right\} \right. \\
&\phantom{\mathcal{H}^{\rm 2-comp} = \frac{\hbar^2 n}{8 m} [} \left. \vphantom{\left\{ \left(\nabla \varphi\right)^2 + \left(\nabla \Delta\varphi\right)^2 \right\}} + 2 (n_1 - n_2) \left(\nabla \varphi\right) \cdot \left(\nabla \Delta\varphi\right) \right] \\
&\phantom{\mathcal{H}^{\rm 2-comp} =} + \frac{(g_1 + g_2) (n_1 + n_2)^2}{4} \\
&\phantom{\mathcal{H}^{\rm 2-comp} =}+ \frac{(g_2 - g_1) (n_1 - n_2)^2}{4} \\
&\phantom{\mathcal{H}^{\rm 2-comp} =}- q_{\rm J} \sqrt{n_1 n_2} \cos\Delta\varphi,
\end{split}
\end{align}
respectively.
By setting $\gamma = \mathrm{const}$ in $\mathcal{H}^{\rm spin-1}$ and $n_1 = n_2 = n/2$, $g_1 = g_2 = \lambda_0$, $q_{\rm J} = q_{\rm Z}$, $\varphi = 2 \phi$, and $\Delta\varphi = 2 \beta$ in $\mathcal{H}^{\rm 2-comp}$, the two Hamiltonian densities become equivalent to each other.
In this sense, we can expect that the same situation can occur in the both systems.
The half-quantized vortices can be expressed as
\begin{align}
\begin{split}
& \psi^{\rm spin-1}_{\rm HQV\pm} = \psi^{\rm spin-1}(\pm \theta/2,\gamma,\theta/2) \\
&\phantom{\psi^{\rm spin-1}_{\rm HQV\pm}} = \frac{1}{2} \sqrt{\frac{n}{2}} \begin{pmatrix} \pm e^{- i \gamma} (e^{i \theta} - 1) \\ \sqrt{2} (e^{i \theta} + 1) \\ \pm e^{i \gamma} (e^{i \theta} - 1) \end{pmatrix}, \\
& \psi^{\rm 2-comp}_{\rm HQV+} = \psi^{\rm 2-comp}(\theta,\theta) = \begin{pmatrix} \sqrt{n_1} e^{i \theta} \\ \sqrt{n_2} \end{pmatrix}, \\
& \psi^{\rm 2-comp}_{\rm HQV-} = \psi^{\rm 2-comp}(\theta,-\theta) = \begin{pmatrix} \sqrt{n_1} \\ \sqrt{n_2} e^{i \theta} \end{pmatrix},
\label{eq:HQV}
\end{split}
\end{align}
where $\theta$ is the angle for the path encircling the vortices.
We emphasize that any half-quantized vortices should have at least one energetically unfavorable point with $\beta = \pi/2$ or $- \pi/2$ for $\mathcal{H}^{\rm spin-1}$ and $\Delta\varphi = \pi$ for $\mathcal{H}^{\rm 2-comp}$ along the path $\theta$, and they are topologically unstable when $q_{\rm Z,J} > 0$.
On the other hand, the integer vortices expressed as
\begin{align}
\begin{split}
  & \psi^{\rm spin-1}_{\rm IV} = \varphi(\beta,\gamma,\theta) = e^{i \theta} \sqrt{\frac{n}{2}} \begin{pmatrix} i e^{- i \gamma} \sin\beta \\ \sqrt{2} \cos\beta \\ i e^{i \gamma} \sin\beta \end{pmatrix}, \\
  & \psi^{\rm 2-comp}_{\rm IV} = \psi^{\rm 2-comp}(2 \theta,\Delta\varphi) \\
  &\phantom{\psi^{\rm 2-comp}_{\rm IV}} = e^{i \theta} \begin{pmatrix} \sqrt{n_1} e^{i \Delta\varphi/2} \\ \sqrt{n_2} e^{- i \Delta\varphi/2} \end{pmatrix},
  \label{eq:IV}
\end{split}
\end{align}
takes arbitrary values of $\beta$ for $\mathcal{H}^{\rm spin-1}$ and $\Delta \varphi$ for $\mathcal{H}^{\rm 2-comp}$ along the path $\theta$.
Because the states $\psi^{\rm spin-1}$ with $\beta = 0$ or $\beta = \pi$ and the $\psi^{\rm 2-comp}$ with $\Delta \varphi = 0$ are energetically favorable, topologically stable integer vortex states are expressed as
\begin{align}
\begin{split}
  & \psi^{\rm spin-1}_{\rm IV} = \sqrt{n} \begin{pmatrix} 0 \\ e^{i \theta} \\0 \end{pmatrix}, \\
  & \psi^{\rm 2-comp}_{\rm IV} = e^{i \theta} \begin{pmatrix} \sqrt{n_1} \\ \sqrt{n_2} \end{pmatrix}.
  \label{eq:IV}
\end{split}
\end{align}

We consider the ansatzes for the vortex molecule $[1,1]_{r_0}$, the vortex-antivortex pairs $\{1,0\}\:\oset{$r_0$}{-}\: \{-1,0\}$ and $\{0,1\}\:\oset{$r_0$}{-}\: \{0,-1\}$, and the molecule-antimolecule pair $[1,1]_\delta\:\oset{$r_0$}{-}\: [-1,-1]_{\delta}$.
In the case of two-component Bose systems, they become
\begin{align}
\begin{split}
& [1,1]_{r_0} \\
&\quad \text{ : } \psi^{\rm 2-comp}(\bar{\theta}^+_{r_0,0}, \bar{\theta}^-_{r_0,0}) = \begin{pmatrix} \sqrt{n_1} e^{i \bar{\theta}_{r_0,0}} \\ \sqrt{n_2} e^{i \bar{\theta}_{-r_0,0}} \end{pmatrix}, \\
& \{1,0\}\:\oset{$r_0$}{-}\: \{-1,0\} \\
&\quad \text{ : } \psi^{\rm 2-comp}(\bar{\theta}^-_{r_0,0}, \bar{\theta}^-_{r_0,0}) = \begin{pmatrix} \sqrt{n_1} e^{i \bar{\theta}^-_{r_0,0}} \\ \sqrt{n_2} \end{pmatrix}, \\
& \{0,1\}\:\oset{$r_0$}{-}\: \{0,-1\} \\
&\quad \text{ : } \psi^{\rm 2-comp}(\bar{\theta}^-_{r_0,0}, -\bar{\theta}^-_{r_0,0}) = \begin{pmatrix} \sqrt{n_1} \\ \sqrt{n_2} e^{i \bar{\theta}^-_{r_0,0}} \end{pmatrix}, \\
& [1,1]_\delta\:\oset{$r_0$}{-}\: [-1,-1]_{\delta} \\
&\quad \text{ : } \psi^{\rm 2-comp}(\bar{\theta}^-_{r_0,\delta} + \bar{\theta}^-_{r_0,-\delta}, \bar{\theta}^-_{r_0,\delta} - \bar{\theta}^-_{r_0,-\delta}) \\
&\quad = \begin{pmatrix} \sqrt{n_1} e^{i \theta^-_{r_0,\delta}} \\ \sqrt{n_2} e^{i \bar{\theta}^-_{r_0,-\delta}} \end{pmatrix},
\label{eq:molecule-ansatz-2comp}
\end{split}
\end{align}
where, $\bar{\theta}_{r_0,\delta} = \tan^{-1}(y-\delta/2)/(x-r_0/2)$ and $\bar{\theta}^\pm_{r_0,\delta} = \bar{\theta}_{r_0,\delta} \pm \bar{\theta}_{-r_0,\delta}$.
The interaction energies between vortices can be calculated by inserting ansatzes [Eq. \eqref{eq:molecule-ansatz-2comp}] into the Hamiltonian [Eq. \eqref{eq:2comp-Hamiltonian}] as
\begin{align}
\begin{split}
& E^{\rm 2-comp}_{\rm int}([1,1]_{r_0}) \sim \varepsilon_{\rm J} r_0, \\
& E^{\rm 2-comp}_{\rm int}(\{1,0\}\:\oset{$r_0$}{-}\: \{-1,0\}) \\
&\quad \sim \frac{2 \pi \hbar^2}{m} n_1 \log(r_0 / \xi_1) + \varepsilon_{\rm J} r_0, \\
& E^{\rm 2-comp}_{\rm int}(\{0,1\}\:\oset{$r_0$}{-}\:\{0,-1\}) \\ &\quad \sim \frac{2 \pi \hbar^2}{m} n_2 \log(r_0 / \xi_2) + \varepsilon_{\rm J} r_0, \\
& E^{\rm 2-comp}_{\rm int}([1,1]_\delta\:\oset{$r_0$}{-}\:[-1,-1]_\delta) \\ &\quad \sim \frac{2 \pi \hbar^2}{m} (n_1 + n_2) \log(r_0 / \xi),
\end{split}
\end{align}
where $\varepsilon_{\rm J} = \gamma \hbar \sqrt{q_{\rm J} (n_1 + n_2) \sqrt{n_1 n_2} / m}$ is the energy density of the kink per unit length with $\gamma = \mathcal{O}(1)$, $\xi_{i} = \hbar / \sqrt{2 m g_1 n_i}$ ($i = 1,2$) is the vortex core size for the $i$-th component, and $\xi = \sqrt{\xi_1 \xi_2}$.
In the same way, the ansatzes in the case of the spin-1 spinor Bose system becomes
\begin{align}
\begin{split}
& [1,1]_{r_0} \\&\quad \text{ : } \psi^{\rm spin-1}\left(\frac{\bar{\theta}^-_{r_0,0}}{2},\gamma,\frac{\bar{\theta}^+_{r_0,0}}{2}\right) \\&\quad = \frac{1}{2} \sqrt{\frac{n}{2}} \begin{pmatrix} e^{- i \gamma} (e^{i \bar{\theta}_{r_0,0}} - e^{i \bar{\theta}_{-r_0,0}}) \\ \sqrt{2} (e^{i \bar{\theta}_{r_0,0}} + e^{i \bar{\theta}_{-r_0,0}}) \\ e^{i \gamma} (e^{i \bar{\theta}_{r_0,0}} - e^{i \bar{\theta}_{-r_0,0}}) \end{pmatrix}, \\
& \{1,0\}\:\oset{$r_0$}{-}\: \{-1,0\} \\&\quad \text{ : } \psi^{\rm spin-1}\left(\frac{\bar{\theta}^-_{r_0,0}}{2}, \gamma, \frac{\bar{\theta}^-_{r_0,0}}{2}\right) \\&\quad = \frac{1}{2} \sqrt{\frac{n}{2}} \begin{pmatrix} e^{- i \gamma} (e^{i \bar{\theta}^-_{r_0,0}} - 1) \\ \sqrt{2} (e^{i \bar{\theta}^-_{r_0,0}} + 1) \\ e^{i \gamma} (e^{i \bar{\theta}^-_{r_0,0}} - 1) \end{pmatrix}, \\
& \{0,1\}\:\oset{$r_0$}{-}\: \{0,-1\} \\&\quad \text{ : } \psi^{\rm spin-1}\left( -\frac{\bar{\theta}^-_{r_0,0}}{2}, \gamma, \frac{\bar{\theta}^-_{r_0,0}}{2}\right) \\&\quad = \frac{1}{2} \sqrt{\frac{n}{2}} \begin{pmatrix} - e^{- i \gamma} (e^{i \bar{\theta}^-_{r_0,0}} - 1) \\ \sqrt{2} (e^{i \bar{\theta}^-_{r_0,0}} + 1) \\ - e^{i \gamma} (e^{i \bar{\theta}^-_{r_0,0}} - 1) \end{pmatrix}, \\
& [1,1]_\delta\:\oset{$r_0$}{-}\: [-1,-1]_{\delta} \\&\quad \text{ : } \psi^{\rm spin-1}\left(\frac{\bar{\theta}^-_{r_0,\delta} - \bar{\theta}^-_{r_0,-\delta}}{2}, \gamma, \frac{\bar{\theta}^-_{r_0,\delta} + \bar{\theta}^-_{r_0,-\delta}}{2}\right) \\
&\quad = \frac{1}{2} \sqrt{\frac{n}{2}} \begin{pmatrix} e^{- i \gamma} (e^{i \bar{\theta}^-_{r_0,\delta}} - e^{i \bar{\theta}^-_{r_0,-\delta}}) \\ \sqrt{2} (e^{i \bar{\theta}^-_{r_0,\delta}} + e^{i \bar{\theta}^-_{r_0,-\delta}}) \\ e^{i \gamma} (e^{i \bar{\theta}^-_{r_0,\delta}} - e^{i \bar{\theta}^-_{r_0,-\delta}}) \end{pmatrix},
\end{split}
\end{align}
and the interaction energies can be calculated as
\begin{align}
\begin{split}
& E^{\rm 2-comp}_{\rm int}([1,1]_{r_0}) \sim \varepsilon_{\rm Z} r_0, \\
& E^{\rm 2-comp}_{\rm int}(\{1,0\}\:\oset{$r_0$}{-}\: \{-1,0\}) \\&\quad \sim \frac{\pi \hbar^2}{m} n \log(r_0 / \zeta) + \varepsilon_{\rm Z} r_0, \\
& E^{\rm 2-comp}_{\rm int}(\{0,1\}\:\oset{$r_0$}{-}\:\{0,-1\}) \\&\quad \sim \frac{\pi \hbar^2}{m} n \log(r_0 / \zeta) + \varepsilon_{\rm Z} r_0, \\
& E^{\rm 2-comp}_{\rm int}([1,1]_\delta\:\oset{$r_0$}{-}\:[-1,-1]_\delta) \sim \frac{2 \pi \hbar^2}{m} n \log(r_0 / \zeta),
\end{split}
\end{align}
where $\varepsilon_{\rm Z} = \gamma \hbar \sqrt{q_{\rm Z} n^2 / (2 m)}$ is the energy density of the kink per unit length, $\zeta = \hbar / \sqrt{m \sqrt{\lambda_0 \lambda_1} n}$ is the vortex core size.
Interaction energies in the two systems are also equivalent to each other by taking $n_1 = n_2 = n/2$ and $q_{\rm J} = q_{\rm Z}$, and we can expect that the BKT transitions of these systems show a similar behavior \cite{Kobayashi}.

\end{document}